\documentclass[a4paper]{jpconf}
\usepackage{graphicx}
\usepackage{iopams}
\usepackage{citesort}

\newcommand{\MYlsim}{~{}_{\textstyle\sim}^{\textstyle <}~}

\begin{document}
\title{Theoretical status of $\bvarepsilon'/\bvarepsilon$}

%\author{Vincenzo Cirigliano,$^1$ Hector Gisbert,${}^2$ Antonio Pich${}^3$
%% \footnote[*]{Speaker} 
%and Antonio Rodr\'{\i}guez-S\'anchez${}^4$}

\author{V. Cirigliano,$^1$ H. Gisbert,${}^2$ A. Pich${}^{3,}$\footnote[4]{Speaker}
and A. Rodr\'{\i}guez-S\'anchez${}^5$}

\address{${}^1$ Theoretical Division, Los Alamos National Laboratory, Los Alamos, NM 87545, USA}

\address{${}^2$  Fakult\"at Physik, TU Dortmund, Otto-Hahn-Str. 4, D-44221 Dortmund, Germany}

\address{${}^3$ IFIC, Universitat de Val\`encia -- CSIC, Parc Cient\'{\i}fic, 
%Catedr\'atico Jos\'e Beltr\'an 2, 
E-46980 Paterna, Spain}

\address{${}^5$ Dep. Astronomy \& Theoretical Physics, Lund U.,
%Department of Astronomy and Theoretical Physics, Lund University, 
S\"{o}lvegatan 14A, SE 223-62 Lund, Sweden}

\ead{cirigliano@lanl.gov, hector.gisbert@tu-dortmund.de, antonio.pich@ific.uv.es, antonio.rodriguez@thep.lu.se}

\begin{abstract}
We briefly overview the historical controversy around Standard Model predictions of $\varepsilon'/\varepsilon$ and clarify the underlying physics. A full update of this important observable is presented, with all known short- and long-distance contributions, including isospin-breaking corrections. The current 
Standard Model prediction, $\mathrm{Re}(\varepsilon'/\varepsilon) = (14\pm 5)\cdot 10^{-4}$ \cite{Gisbert:2017vvj,Cirigliano:2019cpi}, is in excellent agreement with the experimentally measured value.
\end{abstract}

\section{Historical prelude}

The first evidence of CP non-invariance in particle physics was the non-zero value of the ratios
\begin{equation}
\label{eq:eta_def}
\eta_{_{+-}}\,\equiv\, \frac{A(K_L\to\pi^+\pi^-)}{A(K_S\to\pi^+\pi^-)}\,\equiv\, \varepsilon +\varepsilon'\, ,
\qquad\qquad
\eta_{_{00}}\,\equiv\, \frac{A(K_L\to\pi^0\pi^0)}{A(K_S\to\pi^0\pi^0)}\,\equiv\, \varepsilon - 2\,\varepsilon'\, ,
\end{equation}
which mainly originates in the $\Delta S=2$ weak transition between the $K^0$ and $\bar K^0$ states \cite{Cirigliano:2011ny,Tanabashi:2018oca}:
$|\varepsilon|  = | \eta_{_{00}} + 2\, \eta_{_{+-}} |/3 =
(2.228\pm 0.011)\cdot 10^{-3}$.
A tiny difference between the two ratios was reported for the first time in 1988 by the CERN NA31 collaboration \cite{Burkhardt:1988yh}, and later established at the $7.2\sigma$ level with the full data samples of NA31 \cite{Barr:1993rx}, NA48 \cite{Fanti:1999nm,Lai:2001ki,Batley:2002gn} and the Fermilab experiments E731 \cite{Gibbons:1993zq} and KTeV \cite{AlaviHarati:1999xp,AlaviHarati:2002ye,Abouzaid:2010ny}:
\begin{equation}\label{eq:epsp-exp}
\mathrm{Re} \left(\varepsilon'/\varepsilon\right)\; =\;
\frac{1}{3} \left( 1   -\left|\frac{\eta_{_{00}}}{\eta_{_{+-}}}\right|\right) \; =\;
(16.6 \pm 2.3) \cdot  10^{-4}\, .
\end{equation}
This important measurement demonstrated the existence of direct CP violation in the $K^0\to 2\pi$ decay amplitudes, confirming, therefore, the Standard Model (SM) quark-mixing mechanism where CP violation is associated with a $\Delta S=1$ transition.

The pioneering leading-order (LO) estimates of the strong-penguin ($Q_6$) amplitude predicted values of $\varepsilon'/\varepsilon\sim 2\cdot 10^{-3}$ \cite{Hagelin:1983rb}. However, since the large top quark mass enhances the electroweak penguin ($Q_8$) correction that has the opposite sign, 
the first next-to-leading-order (NLO) calculations~ \cite{Buras:1993dy,Buras:1996dq,Bosch:1999wr,Ciuchini:1992tj,Ciuchini:1995cd} found results one order of magnitude smaller than (\ref{eq:epsp-exp}). Larger values around $10^{-3}$ were nevertheless obtained (also at NLO) with model-dependent estimates of non-perturbative hadronic contributions~\cite{Bertolini:1998vd,Hambye:1999yy,Bijnens:2000im,Hambye:2003cy}.

It was soon realised that those calculations claiming small values of $\varepsilon'/\varepsilon$ were missing the important role of the final pion dynamics \cite{Pallante:1999qf,Pallante:2000hk}. The proper inclusion of long-distance contributions with chiral-perturbation-theory ($\chi$PT) techniques gave $\mathrm{Re}(\varepsilon'/\varepsilon) = (17\pm 9)\cdot 10^{-4}$ \cite{Pallante:2001he}. Taking also into account a more refined analysis of isospin-breaking (IB) corrections \cite{Ecker:1999kr,Cirigliano:2003nn,Cirigliano:2003gt}, induced by electromagnetic interactions and the light-quark mass difference, led finally to $\mathrm{Re}(\varepsilon'/\varepsilon) = (19\pm 10)\cdot 10^{-4}$ \cite{Pich:2004ee}, in good agreement with the experimental value but with a rather large uncertainty.

The controversy around $\varepsilon'/\varepsilon$ resurrected in 2015, when the lattice RBC-UKQCD collaboration reported 
$\mathrm{Re}(\varepsilon'/\varepsilon) = (1.38\pm 5.15\pm 4.59)\times 10^{-4}$ \cite{Bai:2015nea,Blum:2015ywa}, $2.1\sigma$ below the experimental measurement. This has triggered a revival of the old naive estimates \cite{Buras:2015xba,Buras:2016fys}, some of them making also use of the lattice data \cite{Buras:2015yba,Kitahara:2016nld,Aebischer:2019mtr}, and a large amount of new-physics (NP) explanations (a list of references is given in Refs.~\cite{Gisbert:2017vvj,Cirigliano:2019cpi}). However, the current lattice simulation needs to be taken with a grain of salt because it fails to reproduce the $I=J=0$ $\pi\pi$ scattering phase shift $\delta_0^0$, which plays a key role in the calculation. The lattice value of $\delta_0^0(m_K)$ is $2.9\sigma$ below the experimental result, a much larger discrepancy than the one quoted for $\varepsilon'/\varepsilon$, and nobody suggests any NP explanation for this phase-shift
%%%$\delta_0^0(m_K)$ 
anomaly. RBC-UKQCD is obviously working hard to fix the problem \cite{Christ:2019}.

In view of the situation, we have performed a complete update of the SM calculation of $\varepsilon'/\varepsilon$, with analytical $\chi$PT techniques, taking into account our current knowledge of all relevant inputs, such as quark masses and non-perturbative low-energy constants (LECs). Our final result \cite{Gisbert:2017vvj,Cirigliano:2019cpi},
\begin{equation}\label{eq:epsp-th}
\mathrm{Re}(\varepsilon'/\varepsilon)\, =\, (14\pm 5)\cdot 10^{-4}\, ,
\end{equation}
is in good agreement with the experimental world average in Eq.~(\ref{eq:epsp-exp}).

\section{Basic dynamical features of $\mathbf{K}\boldsymbol{\to}\bpi\bpi$}

The $K^0\to\pi\pi$ decays can be characterized through the amplitudes ${\cal A}_I = A_I\,\e^{\delta_I}$, where $I=0$ or $I=2$ denote the isospin state of the two final pions ($I=1$ is forbidden by Bose symmetry) and the strong phases $\delta_I$ equal the S-wave $\pi\pi$ scattering phase shifts $\delta_0^I(m_K)$,
in the limit of isospin conservation. Assuming isospin symmetry, a direct fit to the $K\to\pi\pi$ rates gives \cite{Antonelli:2010yf}
\begin{equation}\label{eq:phenonumbers1}
A_0 = (2.704 \pm 0.001)\cdot 10^{-7}\,\mathrm{GeV},
\quad\; 
A_2 = (1.210 \pm 0.002)\cdot 10^{-8}\,\mathrm{GeV},
\quad\;
\delta_0-\delta_2=(47.5 \pm 0.9)^\circ,
\end{equation}
%\mathrm{Re}
where the tiny (CP-odd) imaginary parts of $A_I$ have been neglected. Thus, the kaon data exhibit two important properties:
\begin{enumerate}
\item A spectacular enhancement of the isoscalar amplitude ($\Delta I=\frac{1}{2}$ rule), generated by the strong forces, that suppresses the ratio
$\omega\equiv\mathrm{Re}(A_2)/\mathrm{Re}(A_0) \approx 1/22$ by a factor
sixteen with respect to its naive SM expectation (without QCD) of $1/\sqrt{2}$.
\item A huge phase-shift difference, due to the strong final-state interactions (FSI) of the two pions. Therefore, the amplitudes
$\mathcal{A}_I = 
\mathrm{Dis} (\mathcal{A}_I) + i\, \mathrm{Abs} (\mathcal{A}_I)$ have
large absorptive components $\mathrm{Abs} (\mathcal{A}_I)$, specially the isoscalar one. Neglecting the small CP-odd parts, the known $\pi\pi$ scattering phase shifts at $\sqrt{s}=m_K$,
$\delta_0^0(m_K) = (39.2\pm 1.5)^\circ$ and
$\delta_0^2(m_K) = (-8.5\pm 1.5)^\circ$~\cite{Colangelo:2001df}, 
imply that 
\begin{equation}\label{epsp_abs}
\mathrm{Abs}(\mathcal{A}_0)/\mathrm{Dis}(\mathcal{A}_0) = \tan{\delta_0} \approx 0.82\, ,
\qquad\qquad\quad 
\mathrm{Abs}(\mathcal{A}_2)/\mathrm{Dis}(\mathcal{A}_2) = \tan{\delta_2} \approx - 0.15\, .
%\mathrm{Abs}\left( \mathcal{A}_{0}/\mathcal{A}_{2}\right) \approx 
%\mathrm{Dis}\left( \mathcal{A}_{0}/\mathcal{A}_{2}\right) \, , 
\end{equation}
\end{enumerate}

The short-distance perturbative calculations claiming small SM values of $\varepsilon'/\varepsilon$ are unable to generate the physical phase shifts, {\it i.e.}, they predict $\delta_I = 0$ and, therefore, $\mathrm{Abs} (\mathcal{A}_I) = 0$, failing completely to understand the empirical ratios (\ref{epsp_abs}). Since $A_0 = \sqrt{1 + \tan^2{\delta_0}}\; \mathrm{Dis} (\mathcal{A}_0) \approx 1.3\times \mathrm{Dis} (\mathcal{A}_0)$, missing the absorptive contribution leads to a gross underestimation of the isoscalar amplitude. This unitarity pitfall implies also incorrect predictions for the dispersive components, since they are related by analyticity with the absorptive parts: a large absorptive contribution generates a large dispersive correction that is obviously missed in those naive estimates. 

%%%%%%%%%%%%%%%%%%% Figure %%%%%%%%%%%%%%%%%%%%%%%%%%%%%%%%
\begin{figure}[h]\centering
%\begin{minipage}[c]{14pc}
\includegraphics[width=13pc]{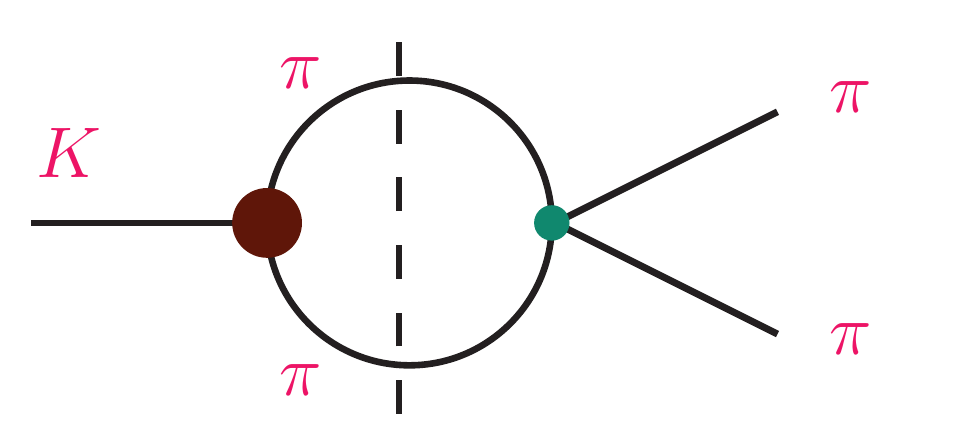}\hspace{3pc}
%\end{minipage}
%\hskip 1.5cm
\begin{minipage}[b]{15pc}
\caption{\label{fig:Abs} Feynman topology generating an absorptive contribution to the $K\to\pi\pi$ amplitudes. The dashed vertical line indicates the corresponding unitarity cut.}
\end{minipage}
\end{figure}
%%%%%%%%%%%%%%%%%%%%%%%%%%%%%%%%%%%%%%%%%%%%%%%%%%%%%%%%%%%

Figure~\ref{fig:Abs} displays the one-loop diagrammatic topology that
generates the absorptive contribution through an on-shell intermediate $\pi\pi$ state. Taking into account the correct chiral structure of the $K\pi\pi$ vertex, the implications of this loop correction are easily computed \cite{Pallante:2001he}:
\begin{equation}\label{eq:AbsLoop}
\Delta \mathcal{A}_0 / \mathcal{A}_0^{\mathrm{tree}}\, =\,
(2 m_K^2-m_\pi^2)\, B_{\mathrm{loop}} +\ldots
\qquad
\Delta \mathcal{A}_2 / \mathcal{A}_2^{\mathrm{tree}}\, =\,
-(m_K^2-2 m_\pi^2)\, B_{\mathrm{loop}} +\ldots
\end{equation}
where the dots stand for contributions from other topologies without absorptive parts, and
\begin{equation}\label{eq:Bloop}
B_{\mathrm{loop}}\, =\, \frac{1}{32\pi^2 F_\pi^2}\,\left\{\sigma_\pi
\left[\log{\left(\frac{1-\sigma_\pi}{1+\sigma_\pi}\right)} + i\pi\right]
+ \log{\left(\frac{\nu_\chi^2}{m_\pi^2}\right)} + 1 \right\}
\end{equation}
with $\sigma_\pi\equiv\sqrt{1-4 m_\pi^2/m_K^2}$.
The two isospin amplitudes get corrections of opposite signs, and the mass-dependent prefactors in Eq.~(\ref{eq:AbsLoop}) make the effect larger by a factor $2.3$ in the isoscalar case.  
The finite one-loop absorptive amplitudes induced by the $i\pi$ term in $B_{\mathrm{loop}}$ are model independent. They represent universal corrections that only depend on the $\pi\pi$ quantum numbers:
\begin{equation}\label{eq:AbsEff}
\mathrm{Abs} (\mathcal{A}_0)/\mathcal{A}_0^{\mathrm{tree}} \, =\, 0.47\, ,
\qquad\qquad
\mathrm{Abs} (\mathcal{A}_2)/\mathcal{A}_2^{\mathrm{tree}} \, =\, -0.21\, . 
\end{equation}

It is worth stressing that these absorptive contributions are present for any effective $K\pi\pi$ vertex in Figure~\ref{fig:Abs}, generating an on-shell intermediate $\pi\pi$ state with the appropriate quantum numbers. Any hypothetical NP contribution at short distances would just modify the denominators in (\ref{eq:AbsEff}), leading to some $\Delta \mathcal{A}_I^{\mathrm{tree}}\sim g_I^{\mathrm{SD}}\mathcal{O}_I$ with some low-energy four-quark operator $\mathcal{O}_I$. Owing to unitarity, the $I=0$ or $I=2$ quantum numbers of this operator determine the same absorptive corrections given in Eq.~(\ref{eq:AbsEff}).\footnote[6]{Unfortunately, most analyses of NP contributions to $K$ decays ignore the presence of absorptive amplitudes.} Moreover, these corrections are identical for the CP-even and CP-odd amplitudes, since they just originate from the real and imaginary parts, respectively, of the short-distance coupling $g_I^{\mathrm{SD}}$ (both in the SM and NP cases).\footnote{The absence of FSI in the CP-odd penguin amplitude has been claimed in Ref.~\cite{Buras:2016fys}, on the basis of an incorrect calculation that violates both chiral symmetry and unitarity.}
The size of these unitarity corrections slightly increases at higher loop orders
\cite{Pallante:2000hk,Pallante:2001he}.

\section{Anatomy of $\bvarepsilon'/\bvarepsilon$}

Direct CP violation appears through the interference between the two isospin amplitudes,
% _{\mathrm{SM}}
\begin{equation}\label{epsp_th}
\mathrm{Re}(\varepsilon'/\varepsilon)\, =\, -\frac{\omega}{\sqrt{2}\, |\varepsilon|}\,\left[\frac{\mathrm{Im} A_0}{\mathrm{Re} A_0} -
\frac{\mathrm{Im} A_2}{\mathrm{Re} A_2}\right]
\, =\,
 -\frac{\omega_+}{\sqrt{2}\, |\varepsilon|}\,\left[\frac{\mathrm{Im} A_0^{(0)}}{\mathrm{Re} A_0^{(0)}}\,\left( 1 -\Omega_{\mathrm{eff}}\right) -
\frac{\mathrm{Im} A_2^{\mathrm{emp}}}{\mathrm{Re} A_2^{(0)}}\right] .
\end{equation}
The observable effect is suppressed by the small value of $\omega$ and is very sensitive to IB contributions \cite{Ecker:1999kr,Cirigliano:2003nn,Cirigliano:2003gt}, parametrized by \cite{Cirigliano:2019cpi}
\begin{equation}\label{eq:omega-eff}
\Omega_{\mathrm{eff}}\, =\, 0.11\pm 0.09 \, ,  
\end{equation}
because small corrections to $A_0$ feed into the small amplitude $A_2$ enhanced by the large factor $1/\omega$. In the right-hand side of Eq.~(\ref{epsp_th}), $\omega_+ = \mathrm{Re}(A_2^+)/\mathrm{Re}(A_0)$ where $A_2^+$ is directly extracted from the $K^+\to\pi^+\pi^0$ rate, 
the $(0)$ superscript denotes the isospin limit, and $A_2^{\mathrm{emp}}$ contains the electromagnetic-penguin contribution to $A_2$ (the remaining contributions are included in $\Omega_{\mathrm{eff}}$).

%%%%%%%%%%%%%%%%%%% Figure %%%%%%%%%%%%%%%%%%%%%%%%%%%%%%%%
\begin{figure}[t]\centering
%\begin{minipage}[c]{14pc}
\includegraphics[width=7pc]{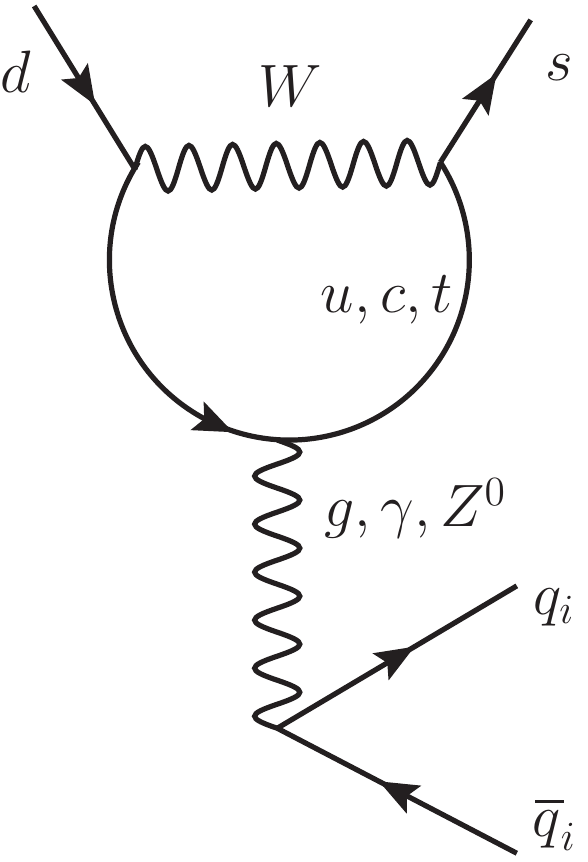}\hspace{3pc}
%\end{minipage}
%\hskip 1.5cm
\begin{minipage}[b]{15pc}
\caption{\label{fig:penguin} Strong and electroweak penguin diagrams, involving the three up-type quarks needed to have CP violation. The CP-odd amplitudes are proportional to the combination of CKM parameters 
$\mathrm{Im} \left(V_{td}^{\phantom{*}} V_{ts}^*\right) = -\mathrm{Im} \left(V_{cd}^{\phantom{*}} V_{cs}^*\right)\approx \eta\lambda^5 A^2$.}
\end{minipage}
\end{figure}
%%%%%%%%%%%%%%%%%%%%%%%%%%%%%%%%%%%%%%%%%%%%%%%%%%%%%%%%%%%

Since $\mathrm{Im} A_2$ is already an IB effect,
$\mathrm{Re} A_2^{(0)}\approx \mathrm{Re} A_2$ can be directly obtained from Eq.~(\ref{eq:phenonumbers1}). The CP-even amplitude $\mathrm{Re} A_0^{(0)}$ is also basically fitted to experimental data \cite{Cirigliano:2019cpi}. Thus, besides the IB parameter $\Omega_{\mathrm{eff}}$, one only needs a theoretical prediction for the CP-odd amplitudes $\mathrm{Im} A_0^{(0)}$ and $\mathrm{Im} A_2^{\mathrm{emp}}$, which to a very good approximation are dominated by the strong ($Q_6$) and electroweak ($Q_8$) penguin operators, respectively: 
\begin{equation}
Q_6 \, =\, -8\,\sum_{q=u,d,s} (\bar s_L q_R)\, (\bar q_R d_L)\, ,
\qquad\qquad
Q_8 \, =\, -12\,\sum_{q=u,d,s} e_q\, (\bar s_L q_R)\, (\bar q_R d_L)\, .
\end{equation}
The hadronic matrix elements of these operators have a sizeable chiral enhancement, due to their scalar/pseudoscalar structure. In the limit of a large number of QCD colours the two colour-singlet quark currents factorize at the hadron level, allowing for an easy determination in terms of their $\chi$PT counterparts. Neglecting the small contributions from all other four-quark operators, one gets
\begin{equation}
    \mathrm{Re}(\varepsilon'/\varepsilon) \, \approx \, 
2.2\cdot 10^{-3}
\left\{ B_6^{(1/2)} \left( 1 - \Omega_{\rm eff} \right) - 0.48\, B_8^{(3/2)}
\right\}  ,\;\label{eq:naiveeps}
\end{equation}
where the factors $B_6^{(1/2)}$ and $B_8^{(3/2)}$ parametrize the deviations of the true matrix elements from their large-$N_C$ approximations.
At $N_C\to\infty$, $B_6^{(1/2)}= B_8^{(3/2)}=1$ and there is a sizeable cancellation between the three terms in (\ref{eq:naiveeps}). Taking $\Omega_{\rm eff}$ from Eq.~(\ref{eq:omega-eff}), one gets $\mathrm{Re}(\varepsilon'/\varepsilon) \approx 0.9 \cdot 10^{-3}$, a factor $2.4$ smaller than the naive $Q_6$ contribution.\footnote{The inputs advocated in Ref.~\cite{Buras:2015yba}, $B_6^{(1/2)}= 0.57$, $B_8^{(3/2)}=0.76$ and $\Omega_{\rm eff} = 0.15$, imply a much larger cancellation leading to $\mathrm{Re}(\varepsilon'/\varepsilon) \approx 2.6 \cdot 10^{-4}$.}
However, this rough estimate does not yet include any chiral loop corrections because they are suppressed by a factor $1/N_C$. In particular, the important logarithmic contributions generating the absorptive components of the amplitudes are totally missed at large $N_C$.
These $\chi$PT corrections increase the $Q_6$ contribution by about 35\% and suppress the $Q_8$ one by 45\% \cite{Pallante:2001he}, destroying the numerical cancellation in Eq.~(\ref{eq:naiveeps}) and bringing back the prediction to larger values.

\section{$\bchi$PT calculation of $\bvarepsilon'/\bvarepsilon$} 

\begin{figure}[th!]\centering
\setlength{\unitlength}{0.52mm} 
\resizebox{1.27\totalheight}{!}{
\begin{picture}(156,121)
\put(0,0){\makebox(156,121){}}
\thicklines
\put(6,104){\makebox(25,7){\textbf{Energy}}}
\put(42,104){\makebox(42,7){\textbf{Fields}}}
\put(102,104){\makebox(52,7){\textbf{Effective Theory}}}
\put(3,103){\line(1,0){156}} {
\put(5,69){\makebox(25,30){$M_W$}}
\put(42,71){\framebox(46,28){
   $\!\!\begin{array}{c} W, Z, \gamma, G_a \\  \tau, \mu, e, \nu_i \\ t, b, c, s, d, u \end{array} $}}
\put(102,69){\makebox(52,30){Standard Model}}

\put(5,34){\makebox(25,20){$\MYlsim m_c$}}
\put(42,36){\framebox(46,19){ 
 $\!\!\begin{array}{c}  \gamma, G_a  \, ;\, \mu ,  e, \nu_i \\ s, d, u \end{array} $}}
\put(102,34){\makebox(52,20){${\cal L}_{\mathrm{QCD}}^{N_f=3}$,
             ${\cal L}_{\mathrm{eff}}^{\Delta S=1}$}}

\put(5,0){\makebox(25,20){$M_K$}}
\put(42,2){\framebox(46,18){ 
 $\!\!\begin{array}{c} \gamma \; ;\; \mu , e, \nu_i  \\  \pi, K,\eta  \end{array} $}}
\put(102,0){\makebox(52,20){$\chi$PT}}
\linethickness{0.3mm}
\put(64,34){\vector(0,-1){12}}
\put(64,69){\vector(0,-1){12}}
\put(69,61){OPE}
\put(69,26){$N_C\to\infty $}}    
\end{picture}
}
\vskip -.5cm\mbox{}
\caption{Evolution from $M_W$ to the kaon mass scale. % $m_K$. %
  \label{fig:eff_th}}
\end{figure}
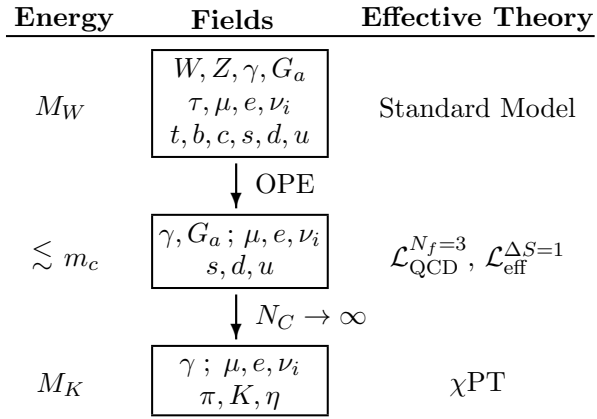

In order to perform a reliable prediction of $\varepsilon'/\varepsilon$ one needs a well-defined effective field theory (EFT) framework, able to control the large logarithmic corrections generated by the presence of widely separated mass scales: $m_\pi < m_K < \nu_\chi \le \mu \ll M_W$. Figure~\ref{fig:eff_th} shows the chain of EFTs needed to describe the relevant physics at the different scales involved.

%%%%%%%%%%%%%%%%%%%%%%%%%%%%%%%%%% Figure %%%%%%%%%%%%%%%%%%%%%%%%%%%%%%%%%%%%%%%%%%%%
%%%%%%%%%%%%%%%%%%%%%%%%%%%%%%%%%%%%%%%%%%%%%%%%%%%%%%%%%%%%%%%%%%%%%%%%%%%%%%%%%%%%%%%%%%

The short-distance QCD logarithmic corrections are very efficiently summed up, all the way down from $M_W$ to $\mu\ge 1$~GeV, with the operator product expansion (OPE). One gets then a $\Delta S=1$ effective Lagrangian
${\cal L}_{\mathrm{eff}}^{\Delta S=1}\sim\sum_i C_i(\mu)\, Q_i(\mu)$,
which is a sum of four-fermion (light-quark) operators $Q_i$, modulated by Wilson coefficients $C_i(\mu)$ that contain all the dynamical information from scales heavier than $\mu$. They are fully known at the NLO \cite{Buras:1991jm,Buras:1992tc,Buras:1992zv,Ciuchini:1993vr} and preliminary NNLO results have been already presented at this conference \cite{cerda-Sevilla:2019,Buras:1999st,Gorbahn:2004my}.

Dynamics at the kaon mass scale can be rigorously described with $\chi$PT, the effective field theory of the QCD Goldstone particles ($\pi,K,\eta$). Using chiral symmetry, one can build the most general effective realization of  ${\cal L}_{\mathrm{eff}}^{\Delta S=1}$ in terms of Goldstone fields, organised as a systematic expansion in powers of momenta over the chiral symmetry breaking scale ($\sim 1$~GeV). Chiral symmetry determines the structure of the $\chi$PT operators with the same symmetry properties as the corresponding four-quark operators $Q_i$, while the short-distance dynamical information is encoded in LECs. The determination of these LECs requires to match the two EFTs in their common region of validity, around 1~GeV. Currently, this can be easily done in the limit $N_C\to\infty$, which turns out to be a very good approximation for $Q_6$ and $Q_8$ because their anomalous dimensions survive the large-$N_C$ limit \cite{Gisbert:2017vvj}.

The great advantage of the $\chi$PT Lagrangian is that it allows for an accurate prediction of the long-distance logarithmic corrections, fulfilling all requirements of unitarity and analyticity. 
In addition to the logarithms generating the absorptive contributions, there are other large chiral logarithms, such as the $\log{(\nu_\chi^2/m_\pi^2)}$ term in Eq.~(\ref{eq:Bloop}), that encode the ultraviolet ($\nu_\chi$) and infrared ($m_\pi$) properties of the EFT and need to be properly taken into account.

Using all this EFT technology, a complete update of the SM prediction of $\varepsilon'/\varepsilon$ has been recently performed in Refs.~\cite{Gisbert:2017vvj,Cirigliano:2019cpi}.
More technical details of this calculation are presented in a separate talk at this conference~\cite{rodriguez-sanchez:2019}. The final result, given before in Eq.~(\ref{eq:epsp-th}), agrees within errors with the experimental world average. Possible improvements in order to further reduce the current theoretical uncertainty have been discussed in Ref.~\cite{Gisbert:2017vvj}.

\section*{Acknowledgements}

%We want to thank the organizers for their effort to make this conference such a successful event. 
This work has been supported in part by the Spanish Government and ERDF funds from the EU Commission [grant FPA2017-84445-P], the Generalitat Valenciana [grant Prometeo/2017/053], 
%%%the Spanish Centro de Excelencia Severo Ochoa Programme [grant SEV-2014-0398], 
the Swedish Research Council [grants 2015-04089  and  2016-05996]  and  the  European  Research Council (ERC) under the EU Horizon 2020 research and innovation programme (grant 668679). The work of H.G. is supported by 
%%%a FPI doctoral contract [BES-2015-073138], funded by the Spanish Ministry of Economy, Industry and Competitiveness and 
the Bundesministerium f\"ur Bildung und Forschung (BMBF). V.C. acknowledges support by the US DOE Office of Nuclear Physics.

\section*{References}
\bibliography{EpsilonpRefs}

\end{document}